\documentclass[twocolumn]{aastex63}

\usepackage{graphicx}
\usepackage[normalem]{ulem}

\def\simgr{\,\hbox{\hbox{$ > $}\kern -0.8em \lower 1.0ex\hbox{$\sim$}}\,}
\def\simle{\,\hbox{\hbox{$ < $}\kern -0.8em \lower 1.0ex\hbox{$\sim$}}\,}

\shortauthors{THORSTENSEN ET AL.}
\shorttitle{Seven Bright Cataclysmics}

\begin{document}

\title{Optical Studies of Seven Bright Southern Cataclysmic
Variable Stars}

\author[0000-0002-4964-4144]{John R. Thorstensen}

\author[0000-0001-5214-9008]{Chase K. Alvarado-Anderson}
\author[0000-0002-5922-4469]{Abigail D. Burrows}
\author[0009-0006-4917-4628]{Rowan M. Goebel-Bain}
\author{David C. Katz}

\affil{Department of Physics and Astronomy,
6127 Wilder Laboratory, Dartmouth College,
Hanover, NH 03755-3528}

\begin{abstract}
We report spectroscopic observations of seven bright southern
cataclysmic variable stars, collected on a single two-week 
observing run using the 1.9-m Radcliffe telescope at the South
African Astronomical Observatory.  
We used radial velocity time series, in some cases in 
combination with other data, to determine or clarify orbital periods 
for five of them, namely ATO J061.1478$-$31.0634, 
BMAM-V547, MGAB-V202, NSV 4202, and V1147 Cen.  
For BMAM-V547, we use data from the Transiting 
Exoplanet Survey Satellite (TESS) to 
corroborate and sharpen the orbital period; the 
TESS data also show a photometric period near 3.93 d, 
likely indicating precession of the accretion disk.
Also, we find a periodic modulation in the radial 
velocities of the SU UMa-type dwarf nova Var Ret2005,
but are unable to specify a unique cycle count.
Finally, we show a spectrum of ASASSN-V J061528.41$-$412007.3 
that appears typical of a luminous novalike variable.
\end{abstract}


\section{INTRODUCTION}

Cataclysmic variables (CVs) are a subclass of mass-exchange binary
stars, in which a white dwarf (WD; the primary) accretes matter from 
a more extended companion (or secondary) that fills its Roche 
critical lobe.  Most commonly, the companion resembles a main-sequence
star, but with differences in detail caused by the complicated
history of mass transfer \citep{knigge06,knigge11}.  Material
transferred from the secondary through the inner Lagrangian point
usually settles into an accretion disk around the white dwarf primary,
and in most CVs the disk dominates the optical luminosity.

The class is diverse; \citet{warner95} gives a comprehensive
review.  If the WD has a strong magnetic field,
it can disrupt the formation of the disk; material instead threads
onto the field and falls down the field lines onto the polar caps,
leading to an AM Herculis star, or `polar', so-called due to the
polarization of their optical light.  If the accretion is very 
slow, the disk can be faint enough for the WD to make a strong
spectral contribution, especially in the ultraviolet; such systems
tend to have very short orbital periods $P_{\rm orb}$.  If $P_{\rm orb}$
is 4-6 hours or longer, the secondary's contribution to the 
combined spectrum often becomes visible.  CVs with $P_{\rm orb} >$
1 d are rare. 

As their name implies, all CVs are variable stars.  Accretion disks
evidently are subject to a limit-cycle instability, leading to dramatic
brightening of typically a few magnitudes, developing over hours
and lasting for several days, during which enough mass is dropped
onto the white dwarf to re-establish the low-density state of the 
disk.  Systems that do this are called {\it dwarf novae} (DNae).
Most known CVs are DNae, and there is an elaborate taxonomy that
describes their outbursts.  Other disk CVs persist in their
high-accretion states; these are the {\it novalike variables} (NLs). 

Spectroscopically, DNe at minimum light show strong Balmer
and \ion{He}{1} lines, greatly broadened by motions in the disk, while
NLs show strong continua.  Some novalikes show almost no emission, while
others show complex line profiles that vary with orbital phase; most
of these are {\it SW Sextantis} stars \citep{thor-swsex91,rodriguez-gil07}.

{\it Population.} All CVs vary, and most call attention to themselves
through their optical variation, though many have been discovered
because of their unusually blue or ultraviolet color, or through
X-ray emission.  The data have become rich and complete enough that
\citet{pala20} created a sample of CVs within 110 pc that they claim
is essentially complete.  However, the pace of discovery remains extremely
high due to the proliferation of high-cadence surveys of sufficient
depth such as ZTF \citep{ZTF}, ASAS-SN \citep{asasn-descr}, 
and ATLAS \citep{tonry18}.  With increasingly-complete samples, it
should be possible to extend Pala's project to much greater depth.
Complete samples are key observables for CV population synthesis 
models such as those of \citet{goliasch15} and \citet{kalomeni16}.

We undertook this study to elucidate the nature of several CVs 
and candidate CVs in the south celestial hemisphere, which remains
somewhat less explored than the north.   We selected our targets from a 
master CV list maintained by the lead author.  Because of time
and aperture constraints, we targeted CVs that remained 
little-studied, in particular objects with unknown or 
uncertain $P_{\rm orb}$.

\section{OBSERVATIONS}

All our observations are from the 1.9 m Radcliffe telescope operated
by the South African Astronomical Observatory.  We used the SpUpNIC 
spectrograph \citep{spupnic} with Grating 6, which covered from 4220 to 
6860 \AA. The 1.$^{\prime\prime}$1 slit yielded a FWHM resolution of $\sim 4.5$
\AA.  Most of our individual exposures were between 8 and 20 min.
We took spectra of a CuAr arc at each new
setting of the telescope, and about once an hour as the telescope tracked.
For our final calibration we used the night-sky airglow lines, 
especially the strong [OI] lines at $\lambda\lambda$ 5577 and 6300,
to adjust the calibration slightly, typically by $\sim 20$ km s$^{-1}$.
The night-sky adjustment failed for a few spectra; for those, we
reverted to the arc calibration.  When the weather was clear, we
observed flux standards in twilight.  From the
scatter in the standard star normalizations -- most
likely caused by seeing variations and the narrow slit -- we
estimate that the absolute calibration is accurate to $\sim 20$ 
per cent, but the relative flux scale should be better than that.

To reduce the data we used a combination of IRAF routines called
from pyraf, and python scripts that made extensive use of 
{\tt astropy} routines.  In particular, we extracted 1-dimensional
spectra from the images using our own implementation
of the variance-weighted extraction algorithm of
\citet{horne86}, as well as the modified wavelength calibration
described earlier.  

We measured radial velocities of the H$\alpha$ emission line --
the strongest emission feature in all these objects -- using
convolution techniques described by \citet{sy80} and \citet{shafter83}.
When a contribution from a late-type star was present, we 
also measured its radial velocity using the {\tt fxcor} task
in IRAF, which implements a cross-correlation technique similar
to \citet{td79}.  For the correlation template, we used the sum 
of 76 spectra of IAU velocity standards, mostly early K stars,
which were individually shifted to zero velocity before summing.  

To search for periods we created an oversampled
grid of test frequencies $\omega$, and at each $\omega$ fit the 
velocities $v(t)$ with a general least-squares sinusoid
$$v(t) = A \sin(\omega t) + B \cos(\omega t) + C,$$
and transformed this to 
$$v(t) = \gamma + K \sin(\omega(t - T_0)).$$
The periodograms we present are derived from these
fits; the quantity plotted as a function of $\omega$ is 
$${1 \over \chi^2} = \left[{1 \over N-3} \sum_{i = 0}^{i = N} \left(v(t) - v_i \over \sigma_i\right)^2\right]^{-1},$$
where the $v_i$ are the $N$ measured velocities, and $\sigma_i$ are
their estimated uncertainties.  The $N-3$ term in the 
denominator arises because at each $\omega$, the three parameters
$K$, $T_0$, and $\gamma$ are adjusted.

When a late-type star was present, we estimated its spectral type
and contribution to the total spectrum using the procedure
described by \citet{peterslate2005}. 

Table \ref{tab:starinfo} lists the stars we observed.  The first column
gives the primary name used in the American Association of Variable
Star Observer's International Variable Star Index (VSX)
\footnote{ at {\tt https://www.aavso.org/vsx/}}.
Some of these objects have multiple designations, generally because they
have appeared in multiple surveys, and VSX lists these designations.
All the objects here are variable, so the $G$ magnitude is only
illustrative.  In the discussion below, we shorten the lengthier 
coordinate-based names.

\begin{deluxetable*}{lrrrrr}
\label{tab:starinfo}
\tablewidth{0pt}
\tablecolumns{6}
\tablecaption{Stars Discussed Here}
\tablecaption{List of Objects}
\tablehead{
\colhead{VSX Name} &
\colhead{$\alpha_{\rm ICRS}$} &
\colhead{$\delta_{\rm ICRS}$} &
\colhead{$G$} &
\colhead{$1/\pi_{\rm DR2}$} &
\colhead{SIMBAD name} \\
\colhead{} &
\colhead{[h:m:s]} &
\colhead{[d:m:s]} &
\colhead{[mag]} &
\colhead{[pc]} &
\colhead{} \\
}  
\startdata
ATO J061.1478$-$31.0634 & 4 04 35.483 & $-$31 03 48.38 & 14.4 & 481(4) & Gaia 19fes \\
Var Ret2005 & 4 11 09.288 & $-$59 11 16.27 & 16.1 & 329(4) & EC 04102-5918 \\
ASASSN-V J061528.41$-$412007.3 & 6 15 28.406 & $-$41 20 07.24 & 13.2 & 636(6) & UCAC4 244-008602 \\
BMAM-V547 & 6 57 33.663 & $-$53 34 22.03 & 14.1 & 1072(17) & Gaia DR3 \ldots \\
MGAB-V202 & 8 18 08.715 & $-$42 34 16.91 & 14.1 & 783(11) & Gaia DR3 \ldots \\
NSV 4202 & 8 39 18.497 & $-$70 32 41.64 & 16.6 & 730(24) & OGLE MC-DN-32 \\
V1147 Cen & 13 00 57.58 & $-$49 12 12.46 & 12.6 & 351(3) & V* V1147 Cen \\
\enddata
\tablecomments{The celestial coordinates and distance estimates are from
the Gaia Data Release 2 \citep{GaiaPaper1, GaiaPaper2}.  Distances are 
inverses of the parallax, without further adjustment.  The SIMBAD designations 
in the final column omit the Gaia numbers for the sake of space. SIMBAD
entries for these objects can be found using coordinates. 
}
\end{deluxetable*}

We list all our radial velocities in Table~\ref{tab:rvels}, and
give parameters of the best-fitting sinusoids in Table~\ref{tab:sinfits}.
The next section discusses the individual stars in greater detail.

\begin{deluxetable*}{llrrrr}
\label{tab:rvels}
\tablewidth{0pt}
\tablecolumns{6}
\tablecaption{Radial Velocities}
\tablehead{
\colhead{Object} &
\colhead{Time\tablenotemark{a}} &
\colhead{$v_{\rm abs}$} &
\colhead{$\sigma$} &
\colhead{$v_{\rm emn}$} &
\colhead{$\sigma$} \\
\colhead{} & 
\colhead{d} &
\colhead{km s$^{-1}$} &
\colhead{km s$^{-1}$} &
\colhead{km s$^{-1}$} &
\colhead{km s$^{-1}$} \\
}
\startdata
ATO J061$-$31 & 59990.2968  & $    0$ & $  14$ & $   48$ & $  11$ \\
ATO J061$-$31 & 59990.3024  & $   35$ & $  11$ & $   39$ & $   9$ \\
ATO J061$-$31 & 59990.3096  & $   53$ & $  10$ & $   34$ & $  10$ \\
ATO J061$-$31 & 59990.3180  & $   73$ & $  11$ & $  -22$ & $  10$ \\
ATO J061$-$31 & 59991.2852  & $   47$ & $  10$ & $   25$ & $   9$ \\
ATO J061$-$31 & 59991.2956  & $   74$ & $  11$ & $  -15$ & $  10$ \\
ATO J061$-$31 & 59991.3061  & $  108$ & $  14$ & $  -21$ & $  10$ \\
\enddata
\tablenotetext{a}{Time of mid-exposure in Barycentric Julian Days,
 minus 2,400,000, referred to UTC.}
\tablecomments{Radial velocities used in this study.  The time argument
is referred to UTC (not TAI) and is the barycentric julian date of mid-exposure
minus 2,400,000., which differs from MJD by 0.5 d.  The full table is 
published as a machine-readable table, and the first few lines are 
shown here to indicate its form and content.}
\end{deluxetable*}

\begin{deluxetable*}{lllrrrr}
\label{tab:sinfits}
\tablewidth{0pt}
\tablecolumns{7}
\tablecaption{Parameters of Sinusoidal Velocity Fits}
\tablehead{
\colhead{Data set} &
\colhead{$T\_0$} &
\colhead{$P$} &
\colhead{$K$} &
\colhead{$\gamma$} &
\colhead{$N$} &
\colhead{$\sigma$} \\
\colhead{} &
\colhead{BJD} &
\colhead{[d]} &
\colhead{km s$^{-1}$}  &
\colhead{km s$^{-1}$}  &
\colhead{} &
\colhead{km s$^{-1}$} \\ 
}  
\startdata
ATO J061-31 abs. & 59994.2295(13) &  0.245282\tablenotemark{a} &  132(5) & $ 39(3)$ & 25 &  11 \\
ATO J061-31 emn. & 59994.363(4) & 0.245282\tablenotemark{a} &  93(10) & $ 4(7)$ & 25 &  23 \\
BMAM-V547 & 59986.376(9) & 0.15536\tablenotemark{b} &  19(7) & $ 12(5)$ & 28 &  17 \\
MGAB-V202 wings & 59988.401(2) & 0.15612(10) &  188(19) & $ 17(12)$ & 76 &  57 \\
NSV-4202 & 59989.553(4) & 0.2839(6) &  80(7) & $-21(5)$ & 31 &  16 \\
V1147 Cen abs. & 59997.3812(13) & 0.4190(5)\tablenotemark{b} &  152(3) & $-34(2)$ & 24 &   8 \\
V1148 Cen emn. & 59997.613(2) & 0.4190 &  134(5) & $-7(4)$ & 24 &  12 \\
\enddata
\tablenotetext{a}{Period held fixed at twice Monard's value.} 
\tablenotetext{b}{Period chosen corresponds to photometric modulation in TESS data.}
\end{deluxetable*}

\section{THE INDIVIDUAL STARS}

\subsection{ATO J061$-$31}

The Catalina light curve \citep{crts} of this bright dwarf nova shows a relatively
steady minimum near $14.4 < V < 14.7$, and a single outburst to 
$V = 12.3$.  It has been followed for some years, but there are
evidently no spectra in the literature.  $P_{\rm orb}$ is not definitively
determined; VSX lists 0.122641(4) d ($\approx$ 2.94 h) from a photometric
modulation at minimum light, attributed to B. Monard\footnote{
The {\tt vsnet-alert} site maintained at Kyoto University retains
an archive of messages about variable stars, mainly CVs. Monthly
digests of the messages can be downloaded from their website,
{http://ooruri.kusastro.kyoto-u.ac.jp/mailman3/postorius/lists/vsnet-alert.ooruri.kusastro.kyoto-u.ac.jp/}.  
Monard's period is relayed by T. Kato in vsnet-alert 23816 (from 2019 December).
}
. The 0.1226-d period would be unusual for a dwarf nova;
CVs near this period, near the long edge of the 
roughly 2- to 3-h `gap' in the 
CV period distribution, tend to be NLs rather than DNae.  
On the other hand, DNae with $P_{\rm orb}$ twice
as long($\sim 5.88$ h) 
often have prominent secondary stars and display two `humps' per orbit
due to the changing aspect of the tidally-elongated secondary.

During our observing run, the target was west of the meridian at evening twilight,
so we could not determine a definitive $P_{\rm orb}$ from our velocities
alone. Our aim instead was to distinguish between candidate periods of 2.94 h 
and 5.88 h.  We obtained 25 exposures totaling 5.9 h, spread over three nights, 
spanning somewhat over 3 h of hour angle.

\begin{figure}
\plotone{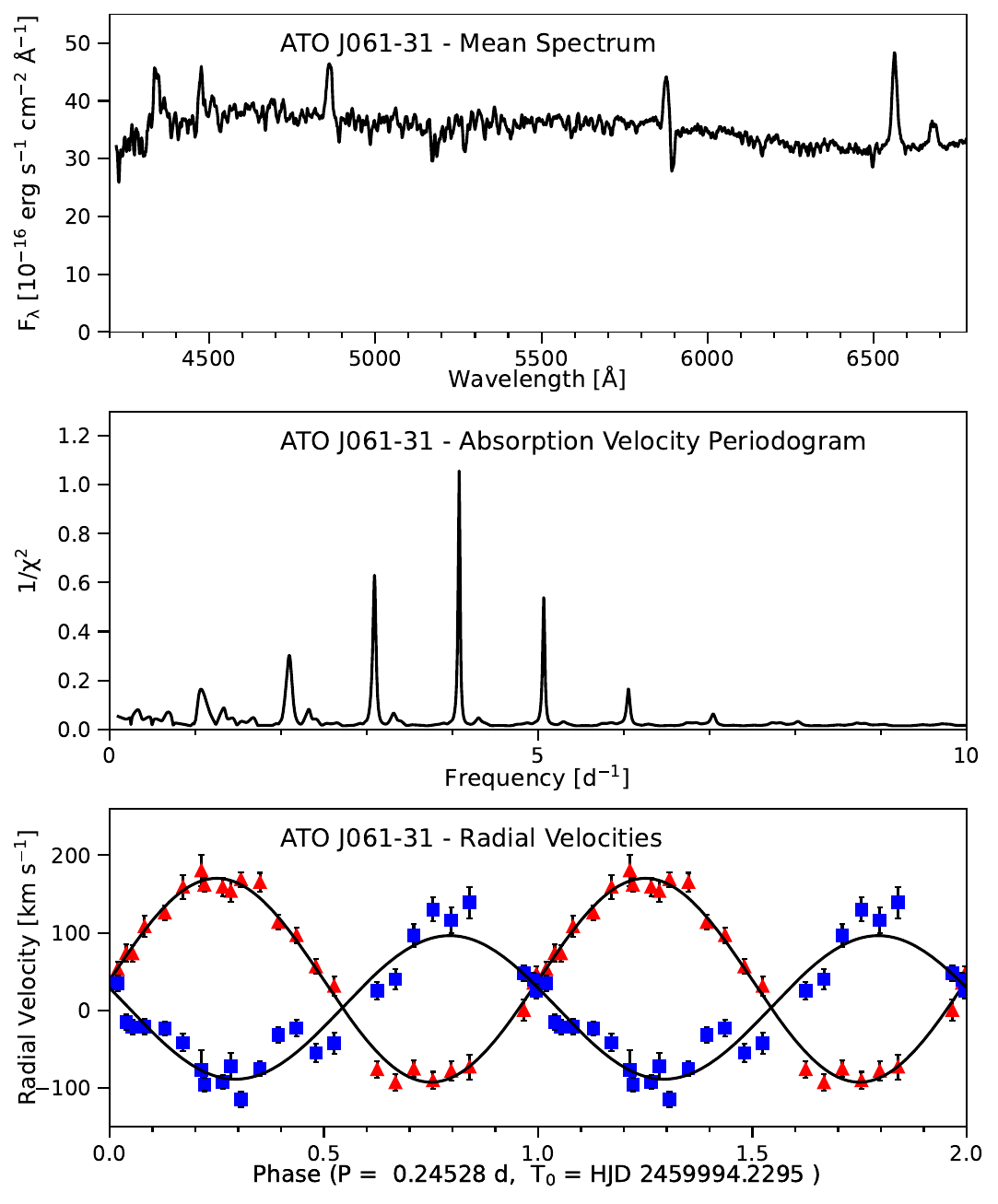}
\caption{ 
{\it Top:} Mean spectrum. {\it Middle:} Periodogram of the absorption 
velocities. Although the sampling was not ideal, the highest peak 
corresponds to the correct period.  {\it Lower:} Absorption (red triangles)
and emission (blue squares) velocities folded on the orbital 
period; the data are repeated over a second cycle to preserve continuity.
The zero of phase is chosen to correspond to the absorption velocity 
increasing through its mean, which should be the inferior conjunction
of the late-type companion.
}
\label{fig:ato061-31montage}
\end{figure}

Fig.~\ref{fig:ato061-31montage} shows the results. The mean spectrum
(top panel) shows multiple absorption features from a late-type 
star in addition to the broadened Balmer and HeI emission
typical of dwarf novae. The middle panel shows the periodogram of the absorption velocities,
which despite the limited sampling, clearly indicates a 5.88-h
period, double the VSX value.  There is no significant modulation
at half this period ($P$ = 2.94 h).  Allowing the period
to vary we find $P = 0.24521(12)$ d, consistent within the 
uncertainties with (double) the more precise Monard period;
we therefore adopt $P_{\rm orb} = 0.245282(8)$ d.
The lower panel shows the folded radial velocities. As expected,
the emission velocities move approximately in antiphase to the 
absorption.

\begin{figure}
\plotone{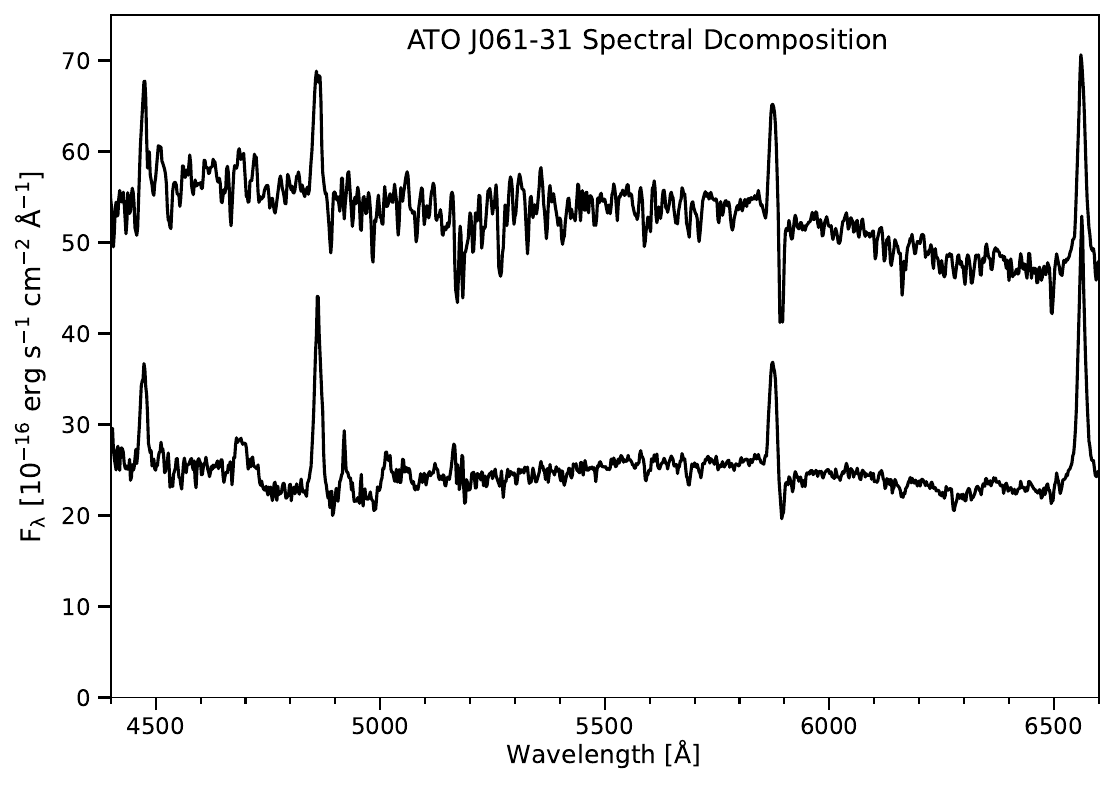}
\caption{ 
The top trace shows the mean spectrum of ATO J061$-$31, averaged after
shifting the individual spectra to the secondary star's rest frame.  
The lower trace shows the same spectrum after subtraction of a 
scaled late-type spectrum (see text).
}
\label{fig:ato061-31decomp}
\end{figure}

The upper trace of Fig.~\ref{fig:ato061-31decomp} shows the mean spectrum, 
and the lower shows it 
after subtraction of a scaled spectrum of the K0.5V-type star, HD124752.
The scaling factor was chosen interactively to best cancel
the late-type features in the difference spectrum.  Our best estimate
of the spectral type is K0-1, with a plausible range from G6 to K4.  
\citet{knigge06} compiled numerous spectral-type estimates for CV
donor stars with known period, and finds that around $P_{\rm orb}$ = 6
hr, the typical spectral type is near M0 (see his Fig.~7).
The secondary in ATO 061-31 therefore appears to be 
significantly warmer than typical.  This might indicate that some 
nuclear evolution has taken place in the secondary. Evolved donor stars can be
much hotter than expected at a given $P_{\rm orb}$ 
\citep{thorfirst1023,thorcss1340,thorasncl,wakamatsu21}.

\begin{figure}
\plotone{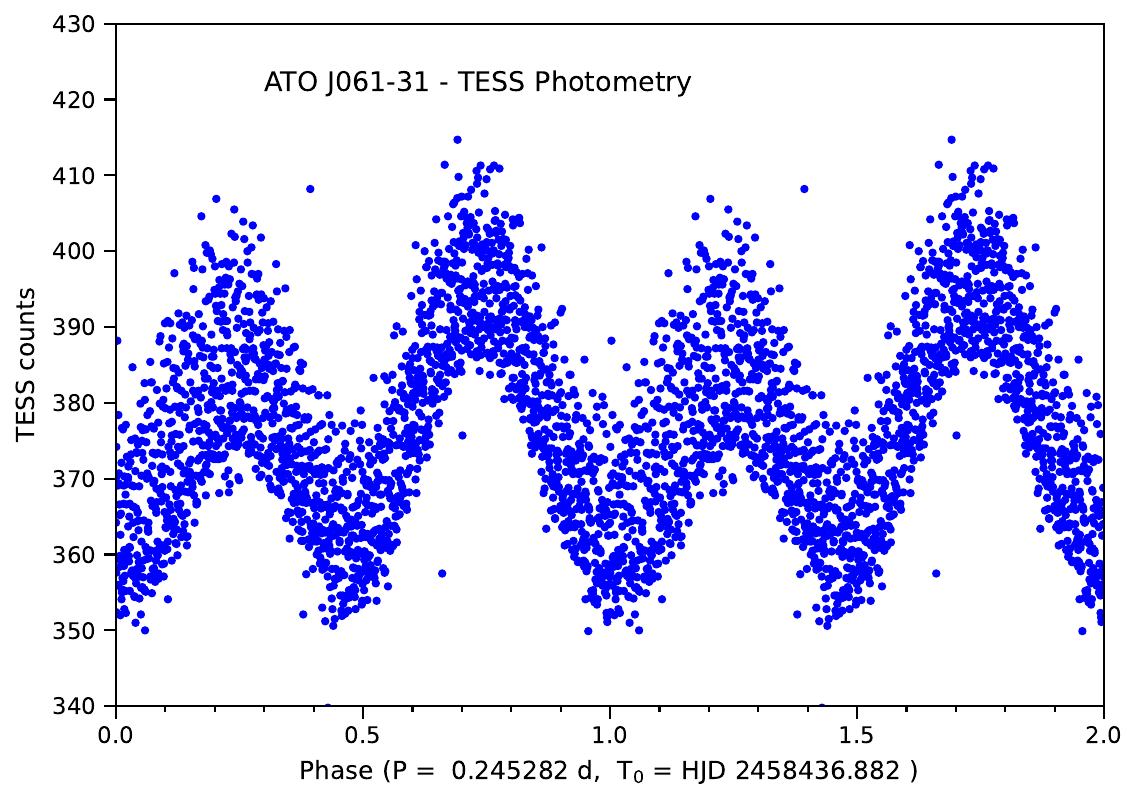}
\caption{ 
TESS data folded on our adopted period.  The period is not known
accurately enough to extend the cycle count back to this epoch,
so the epoch chosen for the fold 
corresponds approximately to one of the two minima.
}
\label{fig:ato061-31tess}
\end{figure}

ATO J061$-$31 was observed by the Transiting Exoplanet Survey Satellite
(TESS) in Sectors 4 and 5, with 1800 sec cadence.  We downloaded
the TESS `PSDCSAP' data using the {\tt lightkurve} python module, edited
out obvious artifacts, and folded the remaining data on the 0.245282-d
period.  The result (Fig.~\ref{fig:ato061-31tess}) shows double-humped 
modulation due to the changing aspect of the tidally-distorted secondary
star.  Note that one maximum appears slightly fainter than the 
other, corroborating once again that the period is 5.88 hours and not half
that.  The TESS data were taken in late 2018, about 4.5 years before our
spectra, and our nominal period (based on doubling the Monard value) 
is not quite precise enough to specify an unambiguous cycle count 
across this gap.  

\subsection{Var Ret2005}

The Gaia alerts-index lists this object as Gaia20cdc, and the Gaia
light curve shows typical quiescent $G$ magnitude
between 16 and 17, but with outbursts to $G \sim 13$ at
intervals ranging from several months to over a year.  
Outbursts were also noted on the vsnet-alert message board;
T. Kato, in vsnet-alert 23341 (2019 July) classified some of
these as superoutbursts, and concluded that the object is 
``almost certainly an SU UMa star.''  However, we were unable
to find mention of any candidate superhump period, nor 
any spectroscopic studies in the literature.

\begin{figure}
\plotone{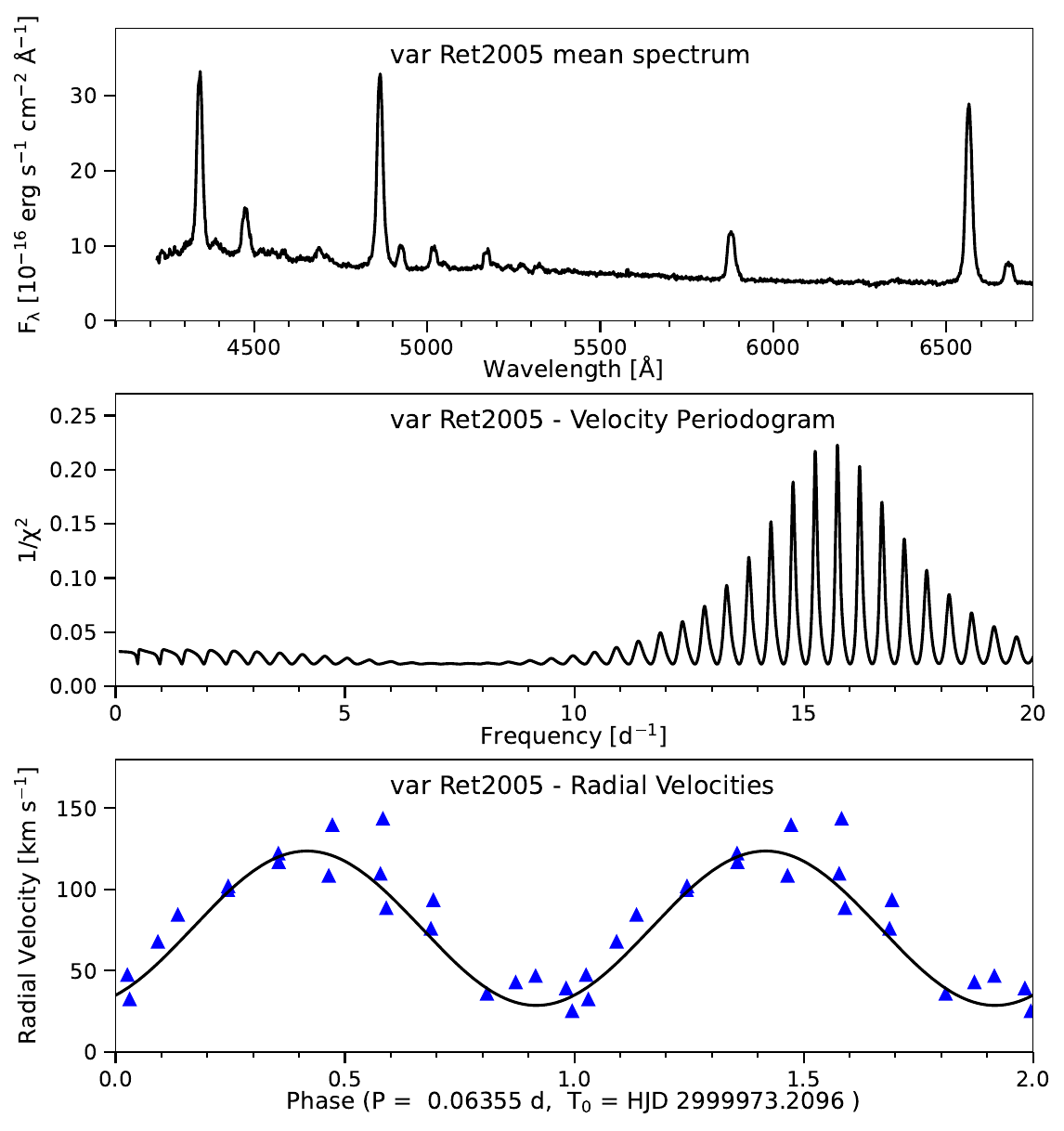}
\caption{ 
{\it Top:} Mean spectrum of Var Ret2005. {\it Middle:} Periodogram of 
the H$\alpha$ emission velocities.   
{\it Lower:} Emission line velocities folded on the best-fitting
period, which is {\it not} determined uniquely.
}
\label{fig:ret2005montage}
\end{figure}

The average of our 20 spectra (Fig.~\ref{fig:ret2005montage}, top)
is typical of 
quiescent short-period dwarf novae (see, e.g., \citealt{thorsuumas20}),
with strong Balmer and HeI emission lines, and no hint
of a late-type companion. 

On the nights we observed we could not obtain a range of
hour angles sufficient to determine an unambiguous
radial velocity period, and weather constrained our 
two visits to be two nights apart.
Consequently, the periodogram
(Fig.~\ref{fig:ret2005montage}, middle) shows
strong aliases spaced by $\Delta f = 1 / (2 {\rm d})$.  
Nonetheless, we constrain the period to the values
shown in Table~\ref{tab:ret2005periods}.  The candidate periods are
shorter than 2 h.  This corroborates Kato's suggested
SU UMa classification, since nearly all dwarf novae in this range
are SU UMa stars.  Assuming the classification is correct,
more complete observations of a superoutburst should reveal
a superhump period, which in turn would resolve the orbital
period ambiguity, since $P_{\rm sh}$ is generally a few 
per cent longer than the orbital period in SU UMa stars 
(see, e.g. \citealt{kato20} and references therein).

\begin{deluxetable}{lrrc}
\label{tab:ret2005periods}
\tablewidth{0pt}
\tablecolumns{4}
\tablecaption{Candidate Periods in Var Ret2005}
\tablehead{
\colhead{Rank} &
\colhead{$P$} & 
\colhead{$1/P$} &
\colhead{$\sigma$} \\
\colhead{} &
\colhead{(d)} & 
\colhead{(d$^{-1}$)} & 
\colhead{(km s$^{-1}$)} \\
}
\startdata
1 & 0.06355 & 15.736 & 11.9 \\
2 & 0.06557 & 15.252 & 12.1 \\
3 & 0.06165 & 16.220 & 12.5 \\
4 & 0.06771 & 14.769 & 13.0 \\
5 & 0.05986 & 16.706 & 13.6 \\
6 & 0.07000 & 14.286 & 14.4 \\
7 & 0.05816 & 17.193 & 15.3 \\
8 & 0.07244 & 13.804 & 16.3 \\
9 & 0.05656 & 17.682 & 17.2 \\
\enddata
\tablecomments{Ranked list of alias periods and
corresponding frequencies from
the Var Ret2005 velocities. The last column gives
the scatter around the best fit at each period.} 
The uncertainties in the individual periods are of
order $5 \times 10^{-5}$ d.
\end{deluxetable}

The lower panel of Fig.~\ref{fig:ret2005montage} 
shows our H$\alpha$ radial velocities
folded at our best period, but readers are cautioned that the 
period chosen is not unambiguous.

\subsection{ASASSN J0615$-$41}

In a 2018 July message on vsnet-chat\footnote{\tt http://ooruri.kusastro.kyoto-u.ac.jp /mailarchive/vsnet-chat/8036},
T. Kato suggested this object is a novalike CV, based on its absolute
magnitude.  
The Catalina Real Time Survey (CRTS; \citealt{crts}) collected 284 magnitudes from 2005 to
2013, which show irregular variation between 12.8 and 13.8, similar
to many NLs.  The object is listed
in the SIMBAD database as a `star'.  We observed the object because
we were unable to find any published spectra.

\begin{figure}
\plotone{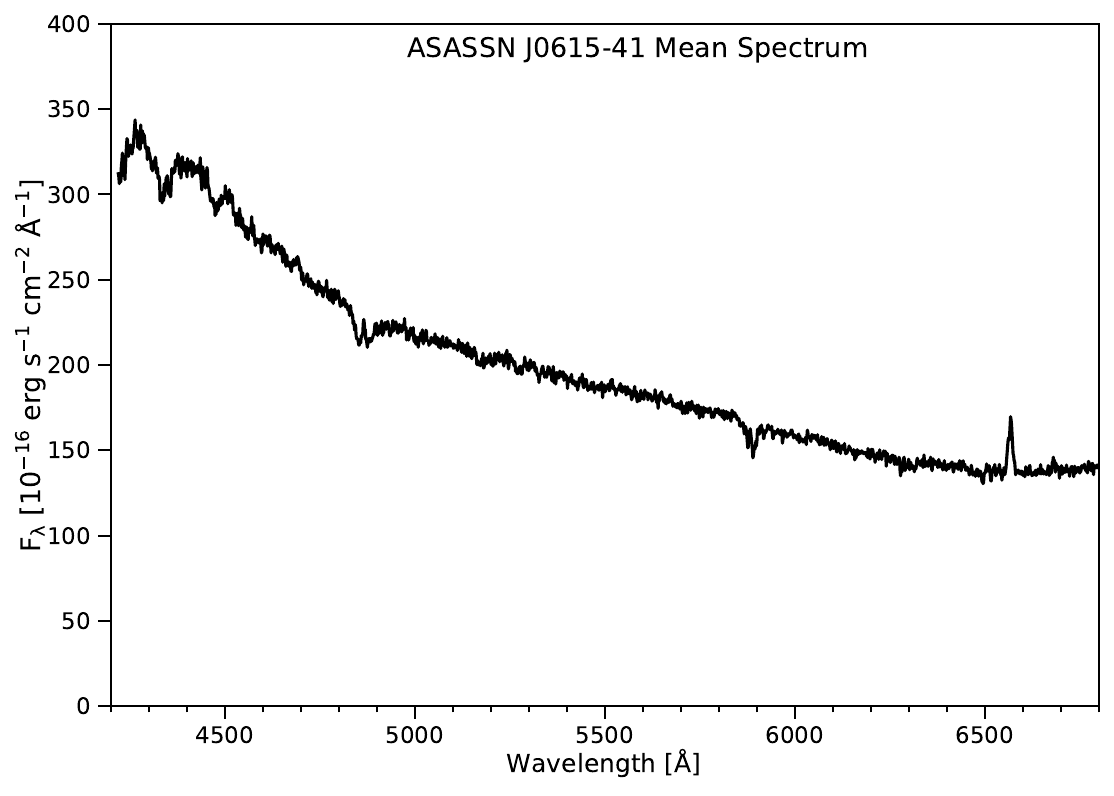}
\caption{
Spectrum of ASASSN 0615$-$41.
}
\label{fig:asn0615-41spec}
\end{figure}

Fig.~\ref{fig:asn0615-41spec} shows the average of two 480-s exposures.  H$\alpha$
emission is present with an emission equivalent width of 3.3 \AA and a FWHM
around 14 \AA .  H$\beta$ is in absorption, with an emission core, and the
higher Balmer lines are entirely in absorption.  The
absorption feature just shortward of $\lambda 5900$ appears to be NaD absorption,
likely interstellar, blended somewhat with weak HeI $\lambda 5876$ absorption.

The spectrum is consistent with a thick-disk, or UX-UMa type, novalike variable
\citep{warner95}.  The variability and spectrum bolster the case
that this is a {\it bona fide} novalike CV.  

\subsection{BMAM-V547}

This object was first noted by 
Mariusz Bajer in archival data\footnote{This and 
MGAB-V202 have apparently not been discussed in the 
literature indexed by SIMBAD and ADS.  Please refer to the 
VSX entries for details.}.
The ASAS-SN light curve shows it varying irregularly around $V = 14.2$,
and more recently fading to about $g = 15.0$, still with 
irregular variations.  It is not classified as a CV in SIMBAD.

\begin{figure}
\plotone{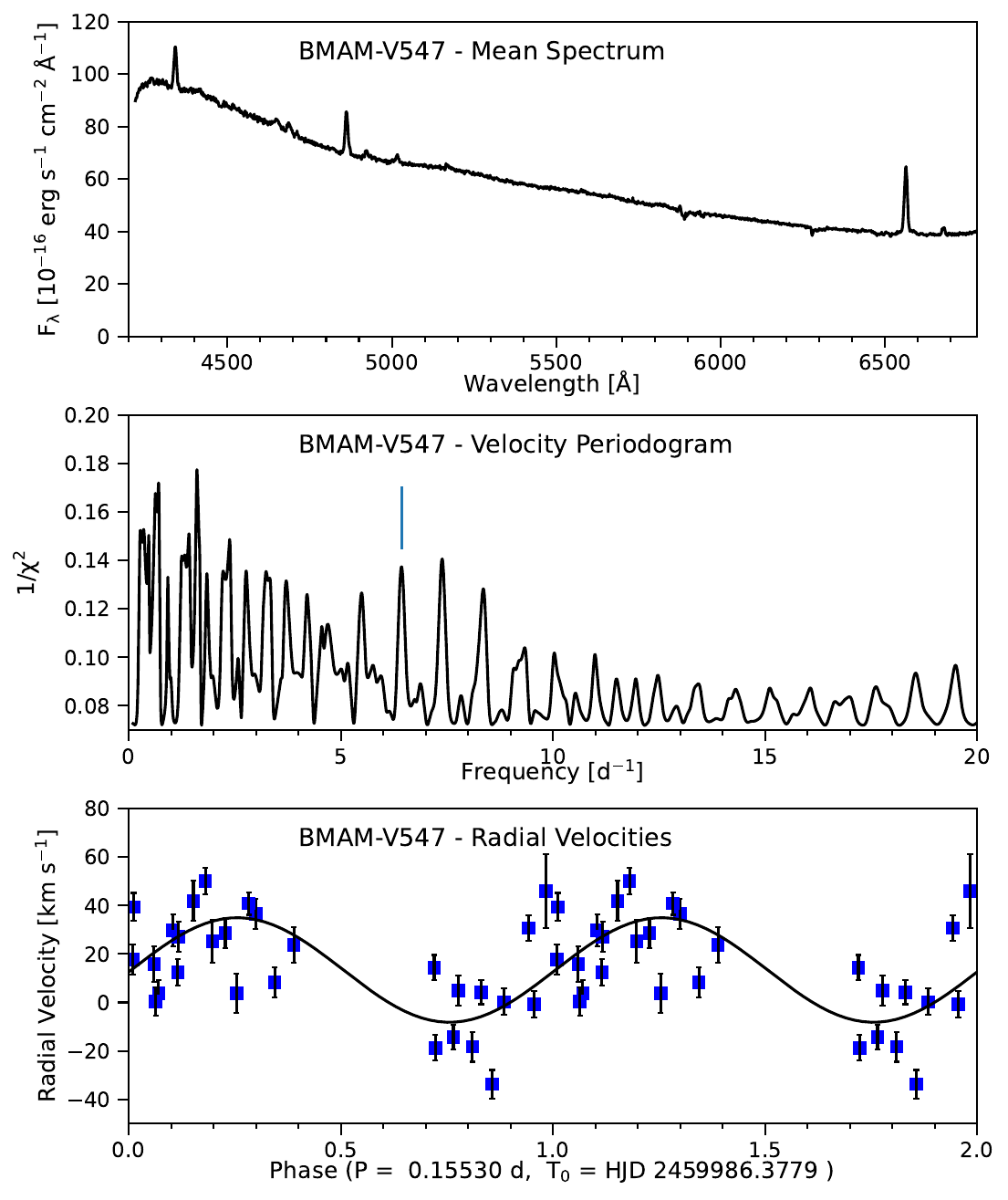}
\caption{
{\it Top:} Mean spectrum of BMAM-V547 from 4.8 h of cumulative exposure.  The 
continuum shape is not precisely determined.
{\it Middle:} Periodogram of H$\alpha$ emission velocities.  The period 
corresponding to the TESS modulation is indicated.
{\it Lower:} H$\alpha$ emission velocities folded on the indicated period.  The
modulation is weak, but apparently significant.
}
\label{fig:bmam547-montage}
\end{figure}

The mean spectrum (top panel of Fig.~\ref{fig:bmam547-montage}) 
shows a strong, blue continuum and weak, narrow emission lines,
typical of a novalike variable.  The amplitude of the emission
radial velocity variations is small, and their periodogram
(middle panel) does not indicate a unique period.  However, one of the 
possible periods, marked with a vertical line in the figure, 
aligns with the photometric period we derive from TESS observations
(see below).  The lower panel shows
the velocities folded on this period, which we identify as
the likely $P_{\rm orb}$.

\begin{deluxetable}{rccc}
\label{tab:bmam547tesssectors}
\tablewidth{0pt}
\tablecolumns{4}
\tablecaption{TESS Observations of BMAM-V547}
\tablehead{
\colhead{Sector} &
\colhead{Start\tablenotemark{a}} &
\colhead{End} &
\colhead{Mean Flux} \\
\colhead{} & 
\colhead{} & 
\colhead{} &
\colhead{(electrons s$^{-1}$)} \\
} 
\startdata
2 & 2018-08-23 &  2018-09-15 &  838 \\
6 & 2018-12-15 &  2019-01-06 &  644 \\
29 & 2020-08-26 &  2020-09-19 &  320 \\
33 & 2020-12-18 &  2021-01-13 &  329 \\
34 & 2021-01-14 &  2021-02-08 &  286 \\
35 & 2021-02-10 &  2021-03-06 &  330 \\
39 & 2021-05-27 &  2021-06-24 &  307 \\
\enddata
\tablenotetext{a}{Dates are the UT of the first and last points
used, in year-month-day form.} 
\end{deluxetable}

\begin{figure}
\plotone{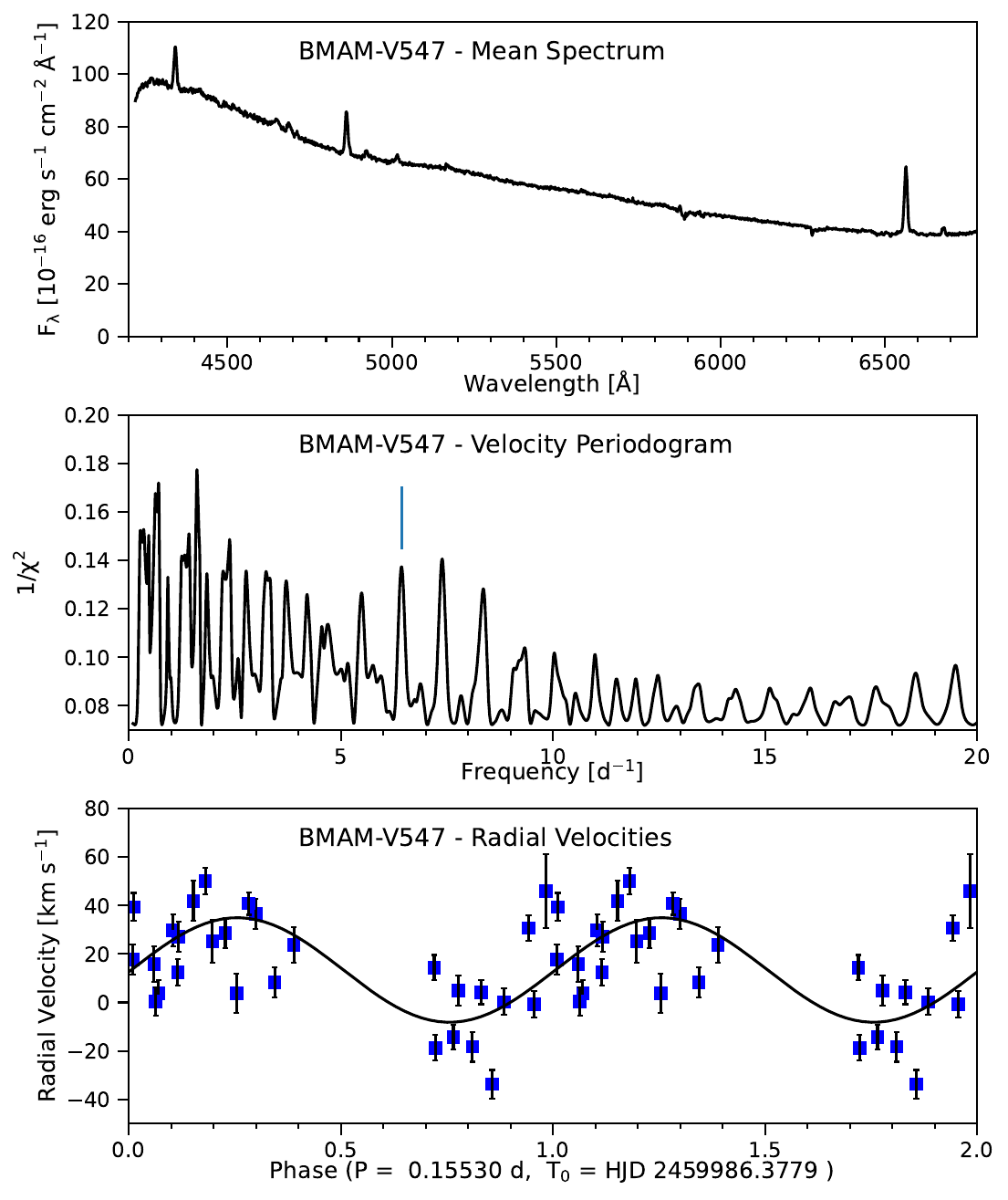}
\caption{
{\it Top:} Lomb-scargle periodogram of the low-state TESS data set (described in 
the text).  The vertical lines indicate the two periods discussed in the text; our 
favored orbital period is near 6.43 d$^{-1}$, and the longer $\sim$3.93-di period, 
likely due to disk precession, is near 0.254 d$^{-1}$.
{\it Middle:} Edited low-state TESS data folded on the likely
orbital period.  Each point is plotted twice for continuity.
{\it Lower panel:} The same data, folded on the $\sim$3.93 d period.  
}
\label{fig:bmam547tessplot}
\end{figure}

We also analyzed TESS observations of this star, which 
are summarized in Table~\ref{tab:bmam547tesssectors}.
We downloaded the PSDCSAP files, edited out apparent
artifacts (and some possible flares, as well, since our aim was
to find periodicities), and computed periodograms using 
the {\tt LombScargle} task from the {\tt astropy} module
{\tt timeseries}.  All the sectors separately showed
very strong modulation near 6.435 cycles d$^{-1}$, 
equivalent to $P = 0.1554$ d, or 3.730 h.  
To explore this, we combined data from four 
sectors in which the mean brightness was 
consistent and relatively low 
-- sectors 29, 33, 35, and 39 -- and searched
this data set for periods (see the top panel of
Fig.~\ref{fig:bmam547tessplot}).  This 
refined the frequency to 6.4365(9) d$^{-1}$,
or $P = 0.15536(2)$ d (near 3.73 hr), where the uncertainty was
estimated by examining light curves folded over a
range of nearby periods.  The modulation apparently maintains coherence over
the 301-day span of the data, which amounts to 
1940 cycles.  The middle panel of Fig.~\ref{fig:bmam547tessplot}
shows the low-state TESS data set folded on this period.

All the TESS data sectors except Sector 2 
(during which the source was brightest) show a second,
weaker modulation (also indicated in Fig.~\ref{fig:bmam547tessplot})
near 0.254 d$^{-1}$, or 3.93 d.   The lower panel of 
Fig.~\ref{fig:bmam547tessplot} shows the low-state
TESS data folded on this much longer period.

The period of the $\sim 3.73$-hr modulation in the 
TESS data is typical of NL orbital periods.
This, together with its coherence and the evidence for radial 
velocity modulation
consistent with the same period, suggests that the 
0.15536-d period is $P_{\rm orb}$, rather than being caused
by some other clock in the system, although given the relatively 
weak velocity modulation we cannot be certain of this.  The spectrum, 
photometric modulation, and velocity modulation are all consistent with 
a novalike variable.

Periods comparable to the 3.93-d period, much longer
than $P_{\rm orb}$ and often called {\it superorbital} periods,
are seen in other novalike CVs (see, e.g., 
\citealt{armstrong13} and references therein). 
These are generally attributed to the precession of
a disk -- either precession of the major axis of an elliptical
disk, or precession of the line of nodes of a tilted disk.  Often, systems
with eccentric or tilted disks also show superhumps, periodic modulations
at frequencies close to the orbital frequency $f_{0}$.
These frequencies are thought to be beats between the precession 
and the orbit, and appear at  $f_0 + f_{\rm p}$ or $f_0 - f{\rm p}$, where 
$f_{\rm p}$ is a disk precessional frequency.  
We do not find these
frequencies in MGAB-V547.  Our favored $P_{\rm orb}$ is flanked by
sidelobes, but these appear to artifacts of the gaps between 
TESS sectors.  In particular, we do not detect noticeable power 
near $f_0 \pm f_{\rm p}$.

\subsection{MGAB-V202}

This object was apparently first identified as a CV candidate by Gabriel
Murawski.  The VSX listing includes a light curve from ASAS-SN showing
irregular variation $13.8 \lesssim V \lesssim 14.4$.  Again, SIMBAD does 
not include a classification as a CV.  

\begin{figure}
\plotone{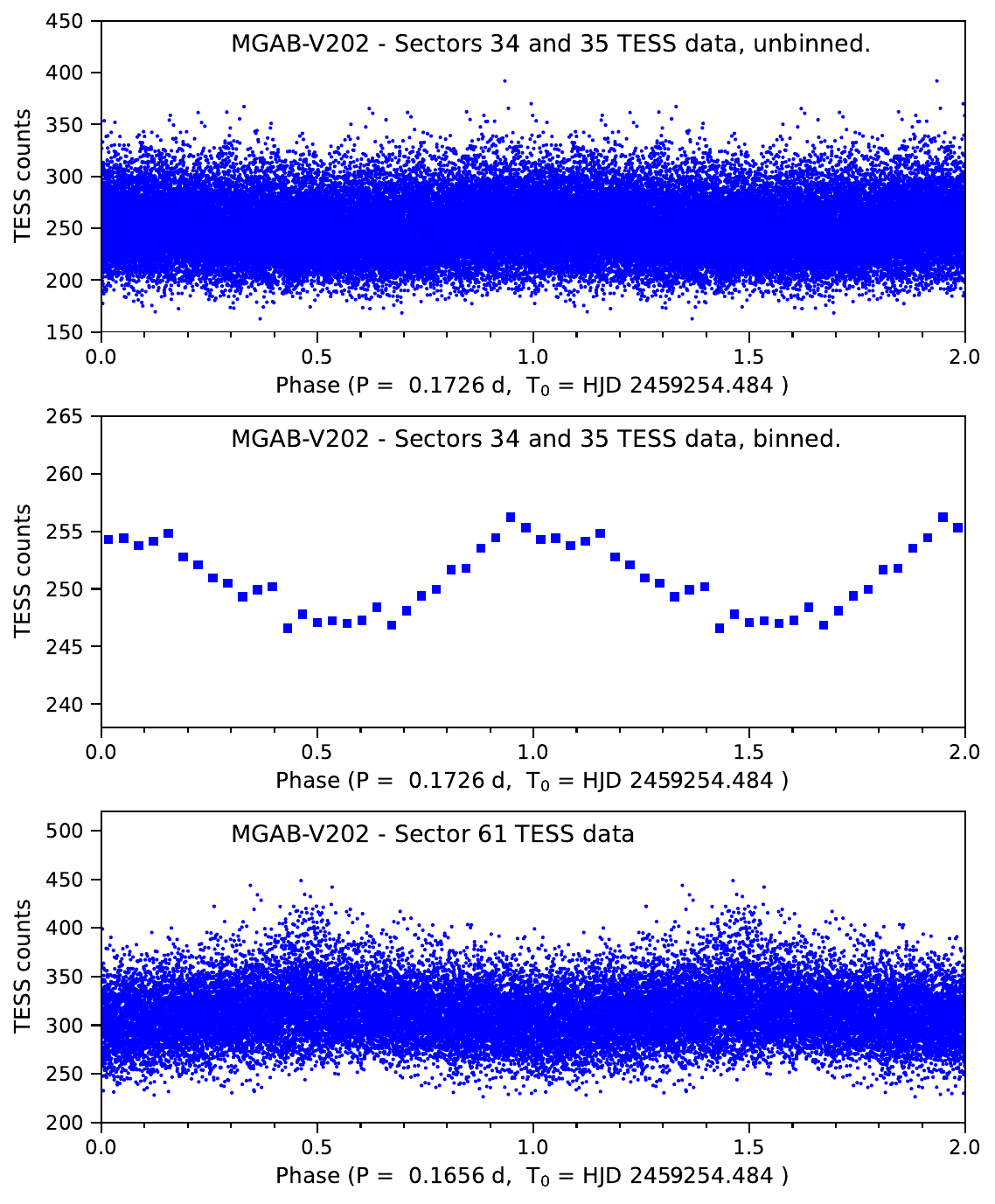}
\caption{
{\it Top:} Unbinned TESS 120-s cadence data from Sectors 34 
and 35 folded on the 0.1726 d period.  No modulation is evident.
{\it Middle:} TESS data from Sectors 34 and 35 averaged in 
phase bins on the 0.1726-d period, to bring out the small
modulation.
{\it Lower:} TESS data from Sector 61, folded on the 0.1656-d
period.  A low-amplitude modulation is evident even though
masked by other variation.
In all these, the zero-point of phase is chosen arbitrarily.
}
\label{fig:mgabtessmontage}
\end{figure}

TESS observed the source in Sectors 34, 35 (2021 February and March,
roughly) and 61 (starting in 2023 January).  Lomb-Scargle periodograms of 
the data from Sectors 34 and 35
both show an apparently significant periodicity near 5.797 cycles d$^{-1}$.
In a simple fold of the data (Fig.~\ref{fig:mgabtessmontage}, top), 
the  modulation is evidently masked by irregular flickering, but 
averaging in phase bins does reveal a low-amplitude modulation (middle panel).
The data from Sector 61, shows a stronger periodicity 
near a {\it different} frequency, 6.037 d$^{-1}$.  This 
modulation is 
discernible in the folded data (lower panel).  In phase-binned averages
(not shown), its maximum and minimum are respectively near 305 and 325 
TESS counts s$^{-1}$.

\begin{figure}
\plotone{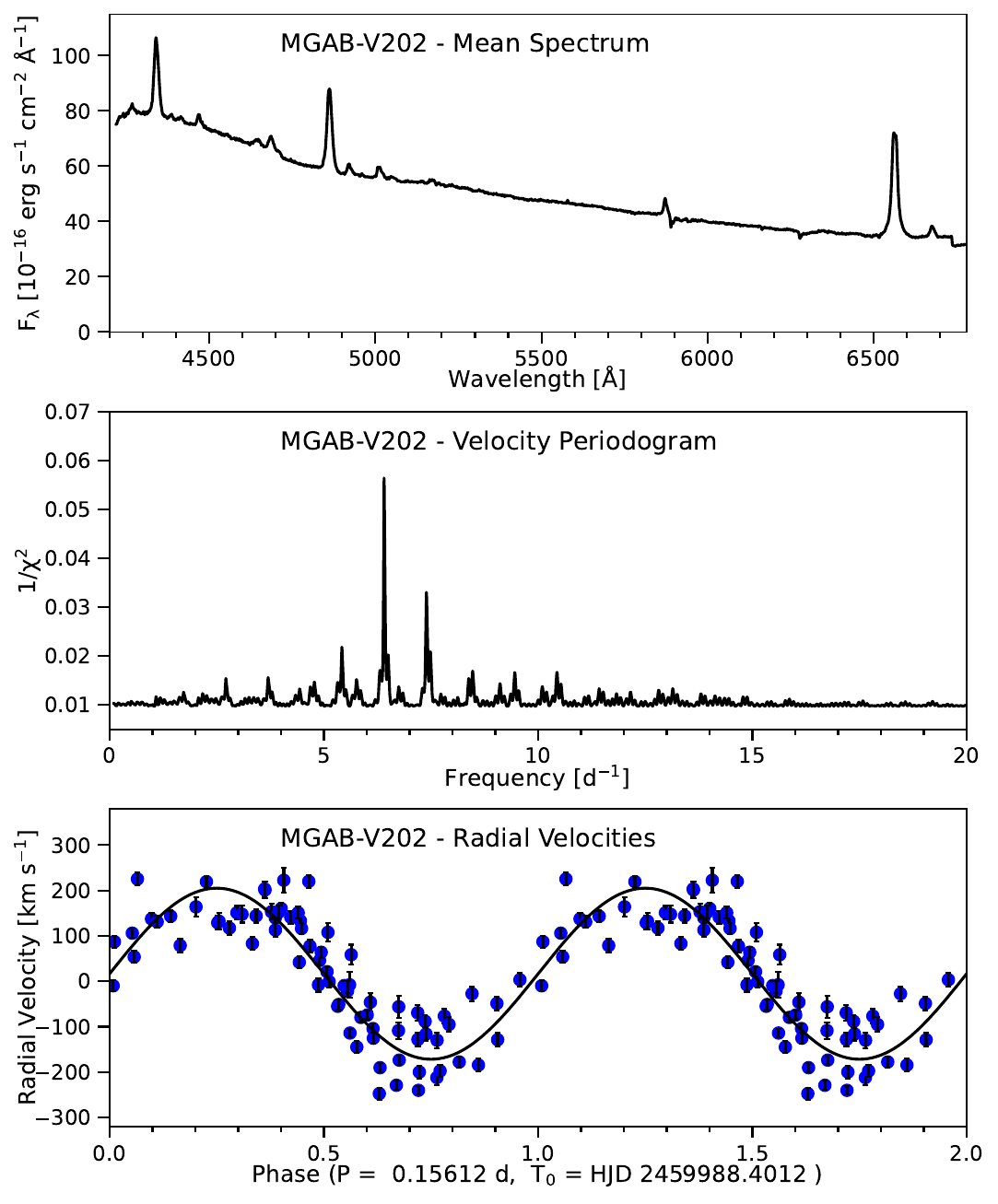}
\caption{
{\it Top:} Mean spectrum of MGAB-V202. 
{\it Middle:} Periodogram of H$\alpha$ line-wing emission velocities.  
{\it Lower:} H$\alpha$ line-wing emission velocities folded on the adopted
period. 
}
\label{fig:mgab202montage}
\end{figure}

We obtained 76 spectra of MGAB-V202, a total of 13.8 h of exposure
time spanning hour angles from $-2.1$ h to $+5.6$ h.  The mean spectrum
(top panel of Fig.~\ref{fig:mgab202montage}) shows relatively strong
Balmer and HeI emission lines on a blue continuum.  For the radial velocities,
we obtained the clearest result using the double-gaussian convolution
with a separation of 42 \AA, which isolated the motion of the rather 
faint wings (or base) of the H$\alpha$ emission line.  This gave the
periodogram shown in the middle panel.  The prominent peak is at 
6.405 cycles d$^{-1}$, or 0.1561 d.  This is, notably, not seen in 
any of the TESS photometry, and it is not a daily alias of any of 
the TESS periods, either.  Thanks to the large span of hour angle,
it is determined without significant ambiguity in the cycle count; a 
1000-trial Monte Carlo simulation of the measurement \citep{tf85} 
returned the correct period every time.  The lower panel shows the 
folded line-wing velocities with the best-fit sinusoid superposed.

\begin{figure}
\plotone{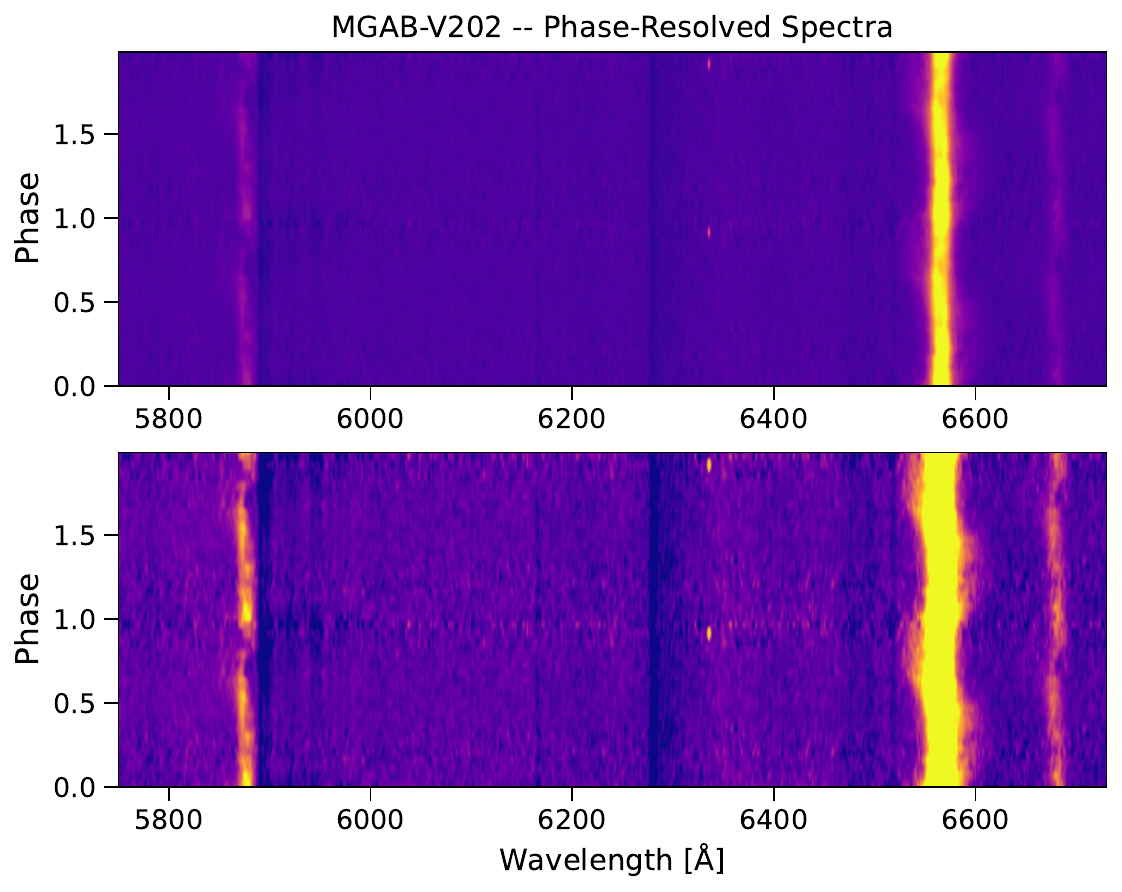}
\caption{
Spectra as a function of phase.  The image was constructed from continuum-divided
spectra.  Each line is a smoothed average of spectra immediately adjacent to the
corresponding phase; there are 100 lines per cycle.  The `stretch' in the upper
is chosen to show the nearly motionless line cores, while the lower panel brings out
the large-amplitude motion of the H$\alpha$ line wings.
}
\label{fig:mgab202trail}
\end{figure}

Fig.~\ref{fig:mgab202trail} displays the spectra as a function of phase
in a two-dimensional image.  The lower panel is `stretched' to show
the large-amplitude motion of the H$\alpha$ line wings.  Also,
the HeI emission lines at $\lambda\lambda$ 5876 and 6678 both show
absorption over part of the phase that appears to drift blueward,
which is a classic symptom of the SW Sex phenomenon \citep{thor-swsex91}.
Faint, large-amplitude Balmer line wings are seen also in 
the novalikes V795 Her \citep{casares96} and LAMOST J204305.95+341340.6
\citep{thorlamost20}.  Based on its spectral appearance, orbital 
period, and detailed spectral behavior, MGAB-V202 is clearly an
SW Sex star.

Both of the photometric periods seen in TESS data
taken $\sim 2$ years apart
are distinct from $P_{\rm orb}$.  
The 0.1725-d period seen in 2021 is 10.5 per cent longer than 
$P_{\rm orb}$, and the stronger 0.1656-d period in early
2023 is 6.1 per cent longer.   As noted earlier,
novalikes in this range
of $P_{\rm orb}$ frequently show {\it superhumps},
either called {\it positive} superhumps, with $P_{\rm sh}$ 
somewhat longer than $P_{\rm orb}$, thought to 
be caused by precession of an eccentric
disk, or {\it negative} superhumps with periods 
shorter than $P_{\rm orb}$, thought to be from apsidal
precession of a tilted disk.
\citet{bruch23} recently studied long-term TESS
light curves of a large sample of novalikes, and found
many examples in which the 
the superhump modulations disappear and/or
change period, as seen here.  The modulations in 
MGAB-V202 appear to be examples of positive superhumps.

\subsection{NSV 4202}

This object was apparently first noticed by \citet{knigge67}.
Sebastian Otero added it to the VSX catalog and classified it as 
a low-amplitude dwarf nova based
on its light curve from ASAS-3.  It was also detected
by the OGLE-III survey \citep{udalski15}.  The ASAS-SN light curve shows
shows a rather flat quiescence near $V = 14.4$, and outbursts to 
$V \sim 12.8$ at irregular intervals of order 100 days, all 
typical of dwarf novae.  We were unable to find any candidate
orbital period in the literature.  

\begin{figure}
\plotone{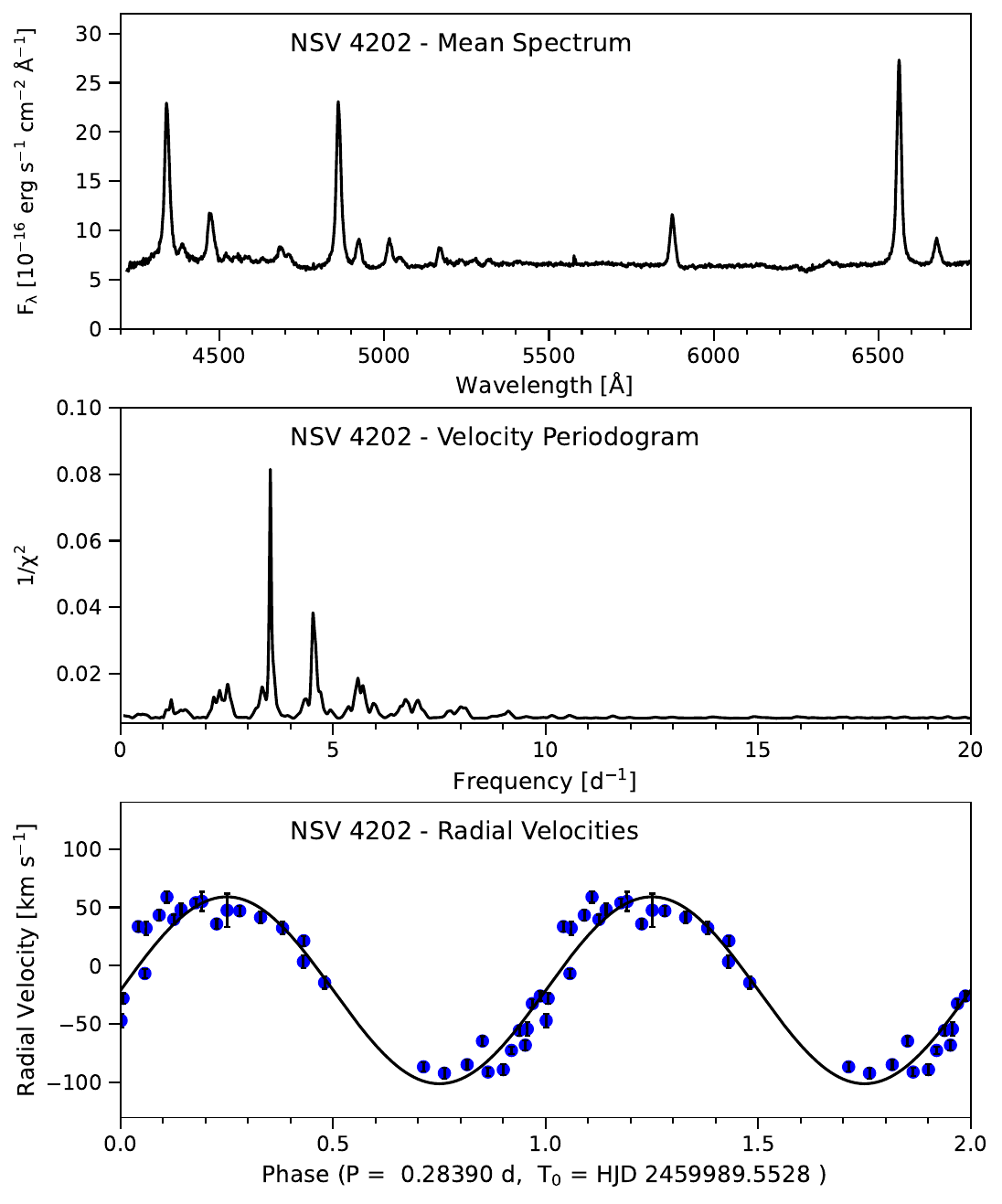}
\caption{
{\it Top:} Mean spectrum of NSV 4202. 
{\it Middle:} Periodogram of H$\alpha$ emission velocities.  
{\it Lower:} H$\alpha$ emission velocities folded on the best period.}
\label{fig:nsv4202montage}
\end{figure}

The mean spectrum (Fig.~\ref{fig:nsv4202montage}, top panel) shows strong Balmer
and HeI emission typical of a dwarf nova at minimum light.  The
H$\alpha$ radial velocities show periodicity at 0.2839(6) d, or
6.81 h. This is 3.52 cycle d$^{-1}$. A daily cycle-count alias
at 4.55 cycle d$^{-1}$, or 5.3 h, is marginally possible but
gives a much poorer fit. The data have good alias discrimination
because of the 6.2-h range of hour angle covered, and because
of the amplitude of the modulation relative to the noise 
(i.e., $K/\sigma$); the Monte Carlo test \citep{tf85}
prefers the stronger period more than 97 per cent of the time.
The lower panel shows velocities folded on the 6.81-h period. 

The period is fairly long for a dwarf nova, but it is somewhat 
surprising that the
mean spectrum shows no contribution from a late-type
secondary.  A secondary contribution is almost always seen
in high signal-to-noise spectra of quiescent dwarf novae with 
periods above six hours or so (see, e.g., the spectra
of ATO 061-31 that were discussed earlier), so we looked for other
evidence to corroborate the period.  Unfortunately, none of 
the synoptic surveys appear
to have sampled this object densely enough to corroborate
our period, and although TESS has observed this deep
southerly location many times, NSV 4202 is $\sim 15$ arcsec
from a significantly brighter star and light curves are
not available.  We also prepared a phase-resolved image
of the rectified spectra, similar to that of MGAB-V202 shown in 
Fig.~\ref{fig:mgab202trail}, but that also showed no 
sign of a secondary star's spectrum.  

\subsection{V1147 Cen}

This object, first noted as a variable star by \citet{luyten35}, 
is apparently the longest-known of those studied here.
It was recognized as a likely U Gem star by \citet{pastukhova07},
who presented an ASAS-3 light curve showing a quiescent level
near 13.5 mag and frequent outbursts to 11.0 mag.  \citet{namelist79}
bestowed the designation V1147 Cen, and listed the type as ``UGSS:''.
The ASAS-SN light curve is 
entirely typical of an active dwarf nova, with outbursts typically
$\sim 40$ days apart.  No detailed study appears to have been 
published, and the orbital period remains unknown.  

TESS light curves are available from Sectors 11 and 37.  Both
show a strong periodicity at 4.767 cycles d$^{-1}$ ($P = 5.035$ h), 
with less power at half that frequency (10.07 h).  

\begin{figure}
\plotone{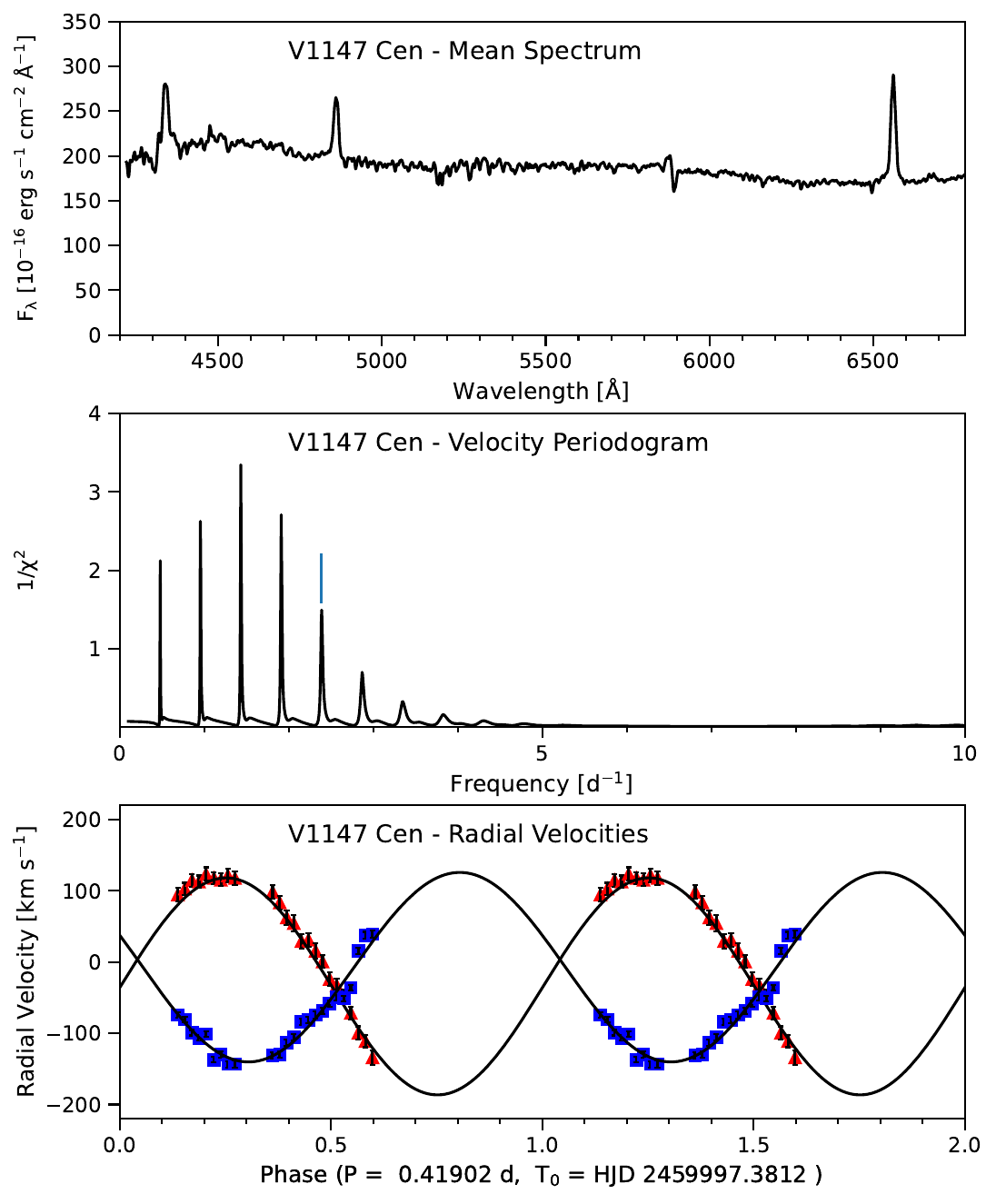}
\caption{
{\it Top:} Mean spectrum of V1147 Cen. 
{\it Middle:} Periodogram of the absorption (cross-correlation) velocities.
The period is not uniquely determined; the vertical tick indicates the
periodicity that is consistent with the TESS light curve.  
{\it Lower:} Absorption (red) and H$\alpha$ emission (blue) velocities 
folded on the adopted period.  The epoch is chosen to correspond to 
inferior conjunction of the secondary star.  The emission
line modulation is approximately, but not exactly, in antiphase to this.}
\label{fig:v1147cenmontage}
\end{figure}

The mean spectrum (Fig.~\ref{fig:v1147cenmontage}, top panel) 
shows typical dwarf nova emission lines and also a contribution
from a late-type star.  We have only 4.0 hours of data covering
4.8 hours of hour angle, so the velocities do not define the
period uniquely, but both the emission and absorption velocities
show a strong, consistent low-frequency modulation 
(Fig.~\ref{fig:v1147cenmontage}, middle panel).  One of the
aliases of this modulation is at $P = 0.4190(5)$ d, or 2.38(3) 
cycle d$^{-1}$, consistent with half the dominant TESS frequency.  
The TESS modulation is clearly due to ellipsoidal variation of 
the secondary, with two humps per orbit.  The lower panel shows 
both emission and absorption velocities folded on the spectroscopic 
orbital period, which amounts to 10.06 h.  

\begin{figure}
\plotone{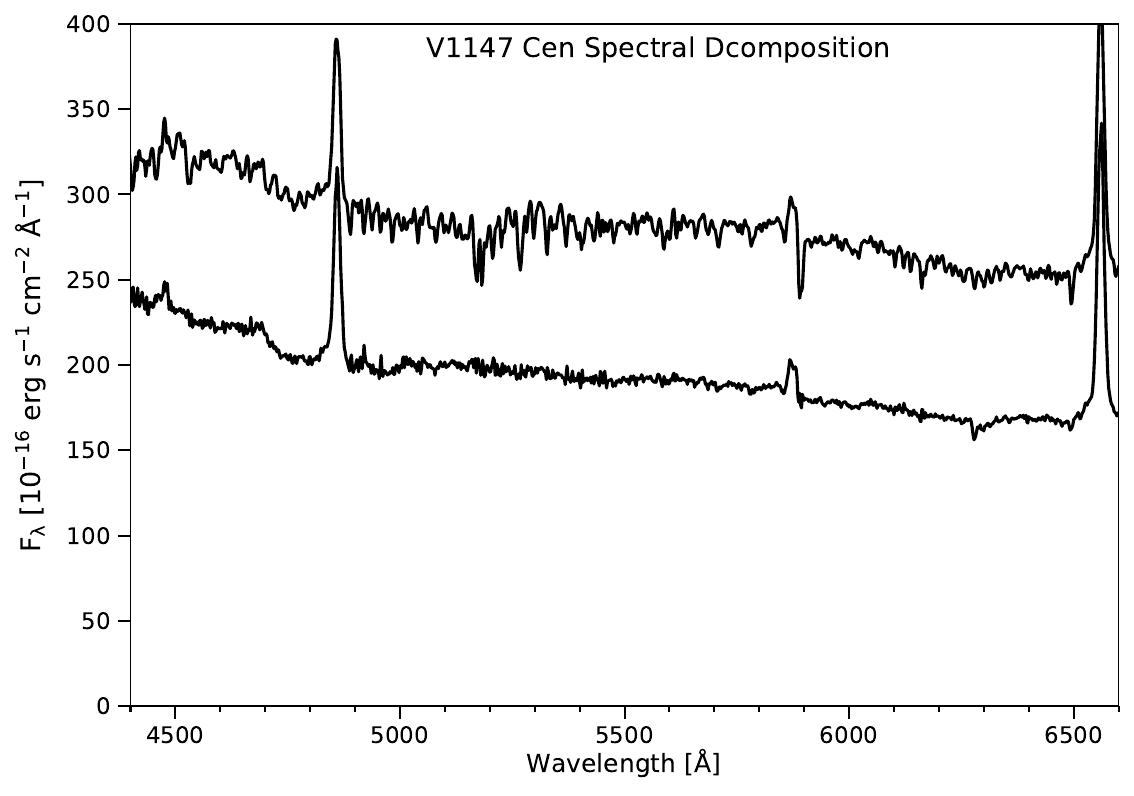}
\caption{
The upper trace shows the average spectrum of V1147 Cen 
in the rest frame of the secondary star, and the lower trace
shows the result of subtracting the K2 star HD109011 from this.
The subtracted spectrum nominally has $V = 14.15.$  The 
apparent bump in the continuum shortward of 4800 \AA\ may
be a calibration artifact.
}
\label{fig:v1147censpecsub}
\end{figure}

The top trace in Fig.~\ref{fig:v1147censpecsub} is the average
flux-calibrated spectrum of V1147 Cen, resampled into the rest
frame of the secondary prior to averaging.  The lower trace
shows the difference between this average and a scaled spectrum 
of the K2V star HD109111.  The secondary features cancel very
well; we found adequate cancellation for types K0 to K3.  
Comparing to other CVs with $P_{\rm orb} \sim 10$ h in
Fig.~7 of \citet{knigge06}, the secondary in V1147 
is cooler than the analytic fit, but similar
to several other examples plotted in his figure and listed in
Knigge's Table 2. 

In summary, we confirm that V1147 Cen is a UGSS star and show that
it has a relatively long
$P_{\rm orb}$.  Had it been more northerly, it would likely have
already been swept up by SDSS, LAMOST, and other surveys and 
attracted more attention.

\section{SUMMARY}

We obtained spectra of selected bright CVs in the southern
hemisphere, with the aim of characterizing them more fully.
The orbital period of a CV is its most fundamental observable,
and for most of our targets we succeeded in measuring
$P_{\rm orb}$.
Our targets, while selected on the basis of tractability, represent
several different subclasses of CVs -- three novalikes, including an
apparent SW Sextantis star, and four dwarf novae, including
two with visible secondary stars and one short-period SU UMa-type
system.  

None of the objects appears grossly atypical, but there are a
few notable findings:  
\begin{itemize}
\item{MGAB-V202 is evidently a new SW Sextantis star.
In TESS photometry it shows two periods clearly different from
$P_{\rm orb}$, neither of which is detected consistently. These may be 
related to disk precession, and further monitoring may be enlightening.}
\item{The TESS photometry of BMAM-V547 shows a clear, persistent modulation
at a period that agrees with one of our possible radial velocity periods. 
In addition, the TESS photometry shows a superorbital period near 3.93 d.}
\item{The secondary star in the dwarf nova ATO J061$-$31
is slightly warmer than expected at its orbital period.}
\item{The dwarf nova NSV 4202 does not show a secondary-star spectrum,
despite its relatively long $P_{\rm orb}$}.   
\end{itemize}

\section{ACKNOWLEDGMENTS}

This paper uses observations made at the South African Astronomical Observatory (SAAO).
We are deeply thankful to the SAAO staff for their warm
hospitality and expert assistance.  Student travel to 
and from the observatory, and
accommodations at SAAO, were underwritten by a generous
donation from Heather and Jay Weed.  The observations 
reported here were taken as part of the Dartmouth College Foreign Study
Program in astronomy; Professors Brian Chaboyer
and Ryan Hickox were essential in arranging, supporting,
and carrying out this program.  

This paper includes data collected with the TESS mission, obtained from the
MAST data archive at the Space Telescope Science Institute (STScI). Funding for
the TESS mission is provided by the NASA Explorer Program. STScI is operated by
the Association of Universities for Research in Astronomy, Inc., under NASA
contract NAS 5–26555

\bibliographystyle{yahapj}
\bibliography{ref}

\end{document}